\documentclass{PoS}
\usepackage{xspace,natbib,color}

\def\etal{{\it et al.}}

\newcommand{\smalltext}[1]{\footnotesize{}#1\normalsize{}}

\definecolor{red}{rgb}{0.7,0.1,0.1}

\newcommand{\refnew}[1]{\ref{#1}\textcolor{red}{}}

\newcommand{\FP}{Fabry-P\'erot\xspace}
\newcommand{\HA}{H$\alpha$\xspace}
\newcommand{\DF}{\textit{DiskFit}\xspace}
\newcommand{\HI}{H\smalltext{I}\xspace}
\newcommand{\LCDM}{$\Lambda$CDM\xspace}
\newcommand{\arcsec}{$''$\xspace}
\newcommand{\ang}{\AA\xspace}

\title{The RINGS Survey: High-Resolution H-alpha Velocity Fields of Nearby Spiral Galaxies with the SALT Fabry-Perot}
\ShortTitle{SALT Fabry-Perot H-alpha Velocity Fields}
\author{\speaker{Carl J. Mitchell}\\
        Rutgers, the State University of New Jersey\\
        E-mail: \email{cmitchell@physics.rutgers.edu}\textcolor{red}{}}
\author{J. A. Sellwood\\
        Rutgers, the State University of New Jersey\\
        E-mail: \email{sellwood@physics.rutgers.edu}\textcolor{red}{}}
\author{T. B. Williams\\
        South African Astronomical Observatory\\
        Rutgers, the State University of New Jersey\\
        E-mail: \email{williams@saao.ac.za}\textcolor{red}{}}
\author{Kristine Spekkens\\
        Royal Military College of Canada\\
        E-mail: \email{kristine.spekkens@rmc.ca}\textcolor{red}{}}
\author{K. Lee-Waddell\\
        Commonwealth Scientific and Industrial Research Organisation\\
        Australia Telescope National Facility\\
        E-mail: \email{Karen.Lee-Waddell@csiro.au}\textcolor{red}{}}
\author{Rachel Kuzio de Naray\\
        Georgia State University\\
        E-mail: \email{kuzio@astro.gsu.edu}\textcolor{red}{}}

\abstract{We have obtained high-spatial-resolution spectrophotometric data on several nearby spiral galaxies with the Southern African Large Telescope (SALT) \FP interferometer on the Robert Stobie Spectrograph (RSS) as a part of the RSS Imaging spectroscopy Nearby Galaxy Survey (RINGS). We have successfully reduced two tracks of \FP data for the galaxy NGC 2280 to produce a velocity field of the \HA line of excited hydrogen. We have modeled these data with the \DF modeling software and found these models to be in excellent agreement both with previous measurements in the literature and with our lower-resolution H\smalltext{I }\small velocity field of the same galaxy. Despite this good agreement, small regions exist where the difference between the \HA and H\smalltext{I }\small velocities is larger than would be expected from typical dispersions. We investigate these regions of high velocity difference and offer possible explanations for their existence.}

\FullConference{SALT Science Conference 2015\\
                 1-5 June 2015\\
                 Stellenbosch Institute of Advanced Study, South Africa}

\begin{document}

\section{Introduction}

The circular speed of gas in a galaxy provides a direct estimate of the central gravitational attraction in that galaxy, and therefore is a probe of that galaxy's mass distribution \citep{Opik22,Robe69,TF77}. Galactic rotation curves are known to flatten at large radii, providing some of the strongest evidence for extended halos of dark matter \citep{Babc39,Rubi80,Bosm81}.

As a cosmological model, Lambda Cold Dark Matter (\LCDM) has been very successful at explaining the origins of large-scale structure in the Universe \citep{Spri06}. However, it is not yet clear whether the small-scale structure of galaxies can be reproduced by \LCDM cosmology. Precise measurements of the distribution of dark matter at small radii within galaxies would provide strong constraints for cosmological models.

A one-dimensional rotation curve $V(R)$ is insufficient to provide such constraints, as the separation of dark and luminous contributions to the rotation curve is degenerate \citep{VanA85,Sack97}. Attempts at resolving this degeneracy have produced vastly different results. Mass models based on theoretical arguments such as spiral-arm multiplicity \citep{SC84,Atha87} and dynamical friction between bars and halos \citep{DS00} favor maximal disk models, in which the luminous disk provides nearly as much of the gravitational force as is permitted by the rotation curve. Hydrodynamical simulations of bar flows have also favored mass models with maximal disks \citep{Wein01}. Measurements of out-of-plane velocity dispersions in disk galaxies have been shown to favor sub-maximal disks, i.e. higher concentrations of dark matter \citep{Bers11}.

In order to address this question, we have designed the RINGS program, for which we are using the \FP interferometer on SALT \citep{Rang08} to obtain 2D velocity fields of the \HA line of excited hydrogen in 19 nearby galaxies. Here we present our \FP observations of one of these galaxies, NGC 2280. We also present \HI 21 cm kinematic observations of this galaxy taken with the Karl G. Jansky Very Large Array (VLA) for the purposes of comparison.

\section{SALT \FP Data}

Our \HA kinematic observations of NGC 2280 were taken on 1 Nov 2011 and 28 Dec 2011 in the medium-resolution mode of the SALT \FP interferometer. On each of these nights, we obtained 25 one-minute exposures of the galaxy, scanning over the \HA line of excited hydrogen from 6565.5 \ang to 6643.5 \ang in 2 \ang steps. The seeing on these nights was approximately 1.75\arcsec and 2\arcsec respectively. The scale of our images after $4\times4$ pixel binning is 0.5\arcsec$/\textrm{pix}$.

The full details of our \FP data reduction are published in \citet{Mitc15}. These include photometric calibrations with images from the CTIO 0.9 m telescope \citep{RINGSphot}, wavelength calibrations, night-sky subtraction, ghost subtraction, and image alignment.

Once the images are aligned and a wavelength solution has been found, choosing a single pixel and measuring the image intensity at that pixel across all of our images yields a coarse spectrum at that point. We have fitted these coarse spectra with Voigt profiles wherever the intensity of the line is strong enough to do so. To improve the signal-to-noise ratio of these fits, we have performed an additional 9$\times$9 (4.5\arcsec $\times$ 4.5\arcsec) spatial binning of the pixels in each individual \FP image. At the distance to NGC 2280, this angular scale corresponds to $\sim 500$ pc.

\begin{figure}

\setbox1=\vbox{{\hsize=.45\hsize \par
    \includegraphics[width=\hsize]{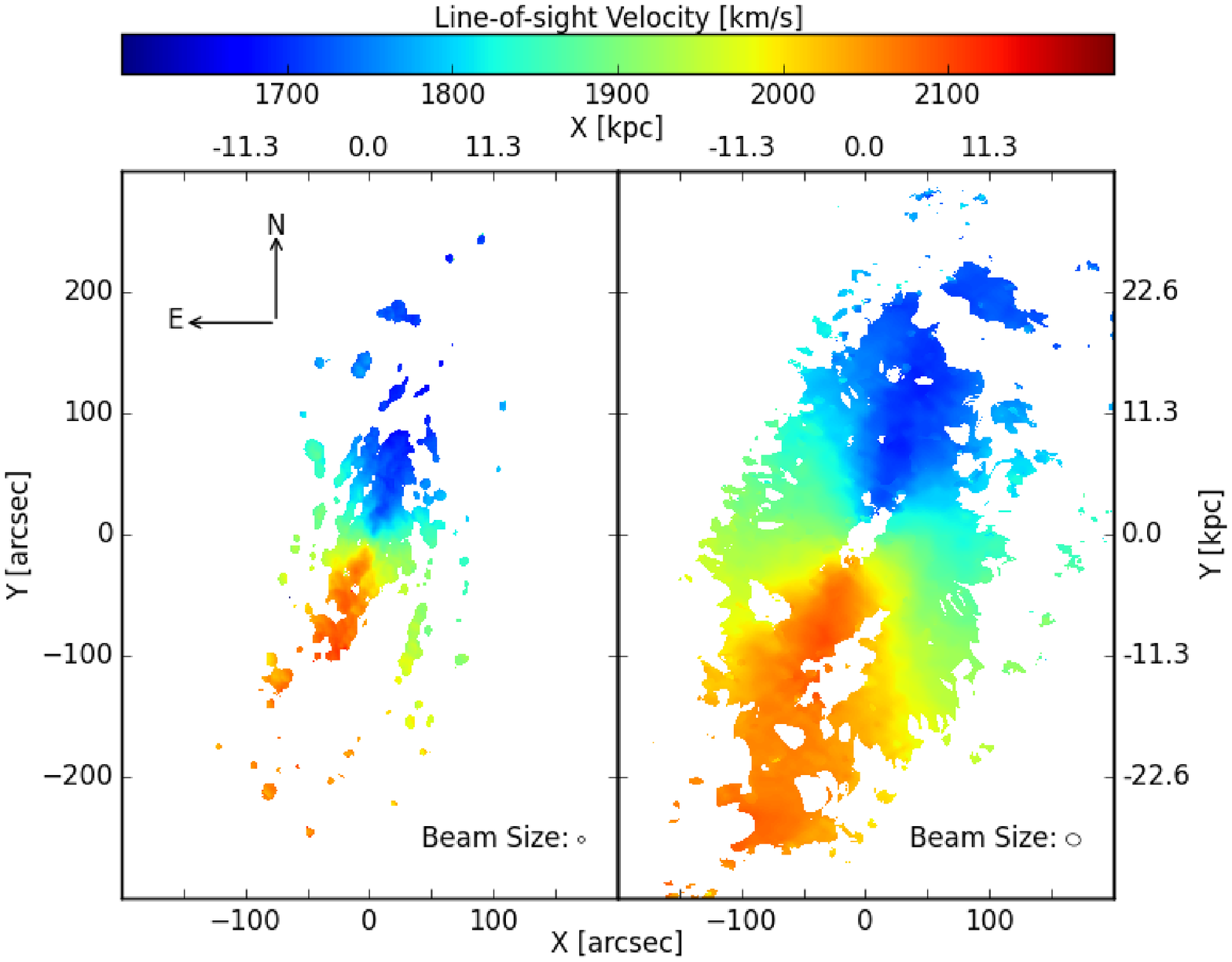}
    \caption{Our \HA (left) and \HI (right) velocity fields of NGC 2280, presented on the same spatial scale. Note the significant differences in spatial resolution, spatial extent, and filling fraction.}\label{fig:velmaps}
\vskip0cm}}

\setbox2=\vbox{{\hsize=.45\hsize 
    \includegraphics[width=\hsize]{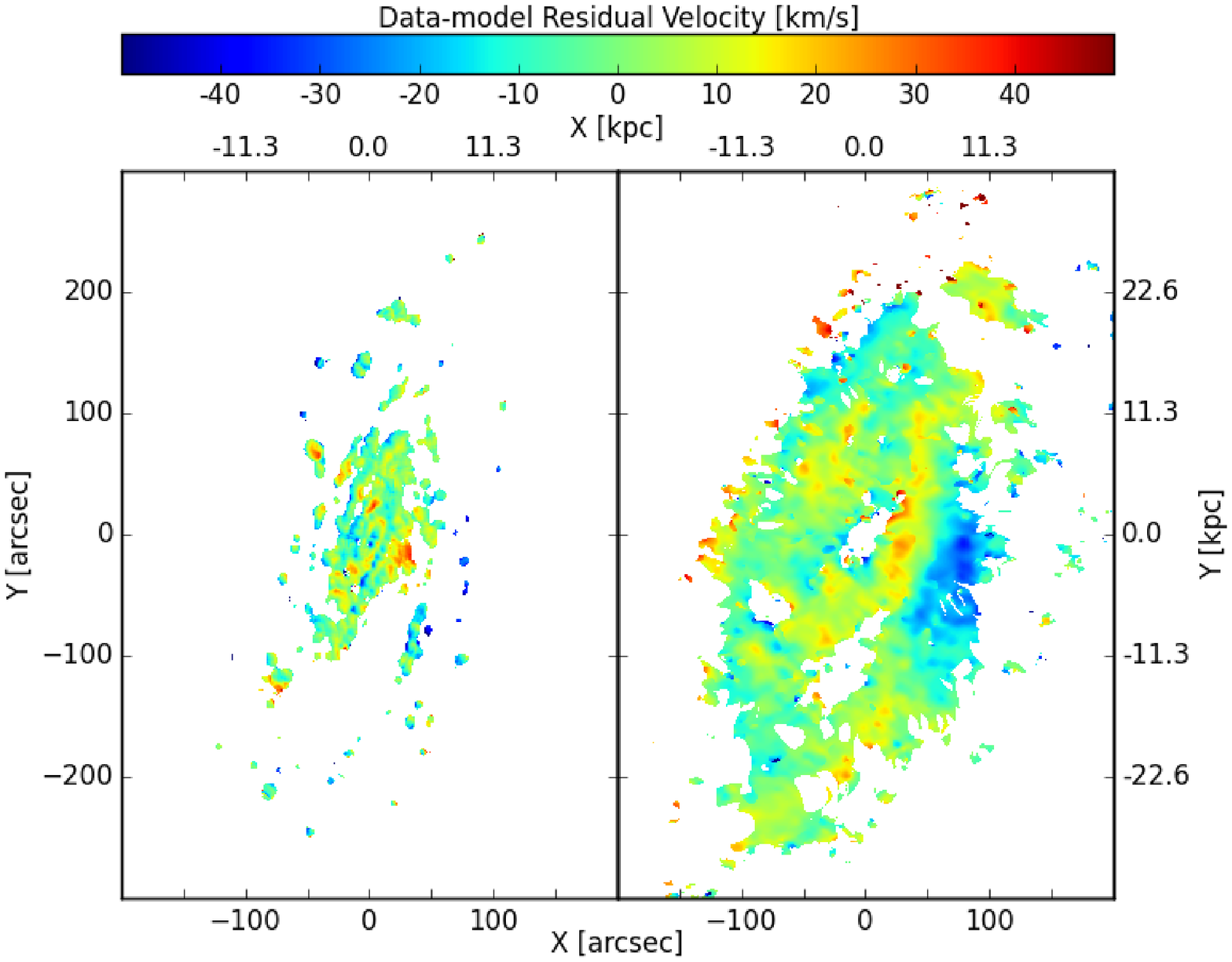}
    \caption{Residual maps to our best-fitting \DF models to our \HA (left) and \HI (right) velocity maps.}\label{fig:resmaps} \vskip 1.05cm}}

\hbox to \hsize{\box1 \hfil \box2}

\end{figure}

The left panel of figure \refnew{fig:velmaps} shows the resulting line-of-sight \HA velocity map.

\section{Kinematic Models}

We have used the \DF\footnote{\DF is publicly available for download at http://www.physics.rutgers.edu/\textasciitilde spekkens/diskfit/} software of \citet{Spek07} and \citet{SeZS10} to fit idealized disk models to our velocity map. The \DF software fits a single tilted disk to the input velocity and uncertainty maps. The fitted parameters are the position of the center, the inclination, the position angle, the systemic velocity, and the circular speed at any number of specified radii. The software is also capable of fitting for asymmetries such as bars and warps, though we find that allowing such flexibility is unnecessary for NGC 2280; such non-axisymmetric models do not significantly improve the fits.

In modeling these data, we have added an additional 8 km/s in quadrature to our uncertainty map to account for the turbulent motions typical of the interstellar medium \citep{Gunn79}. After accounting for this additional 8 km/s, the reduced chi-squared value for our best-fitting model is 2.33, indicating a poor formal fit to the data. The left panel of figure \refnew{fig:resmaps} shows a map of data-minus-model residuals for our best-fitting model. This residual map shows large regions (several effective beams across) of correlated residuals with magnitudes larger than our expected uncertainty. This suggests that the kinematic structure of NGC 2280 is more complicated than is permitted by our axisymmetric model. For example, our models do not include the kinds of streaming motions associated with spiral arms.

\begin{figure}

\setbox1=\vbox{{\hsize=.45\hsize
    \includegraphics[width=\hsize]{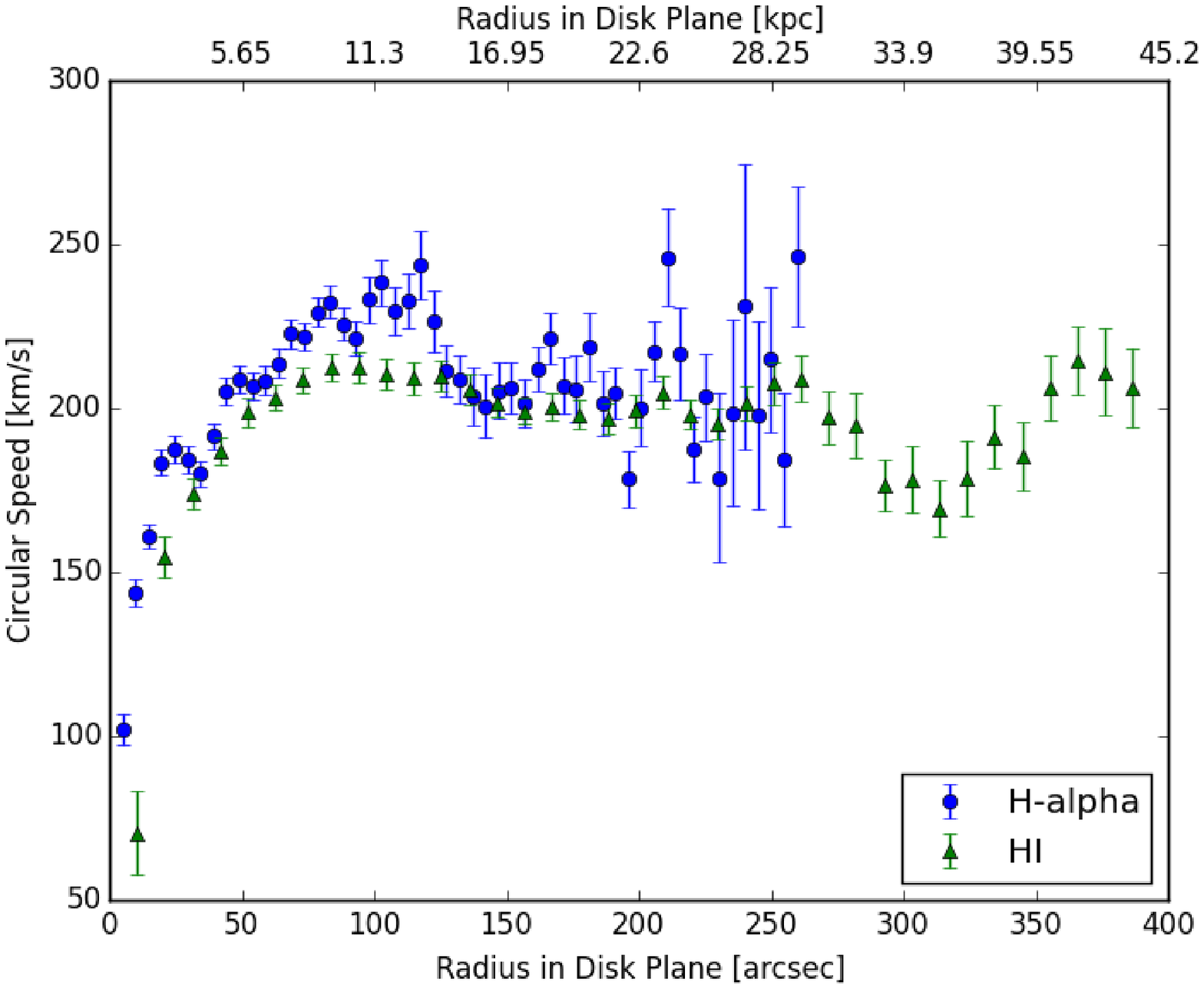}
    \caption{Rotation curves produced from our best-fitting \DF models to our \HA and \HI velocity maps.}\label{fig:rotcurves}
\vskip2.1cm}}

\setbox2=\vbox{{\hsize=.45\hsize 
    \includegraphics[width=\hsize]{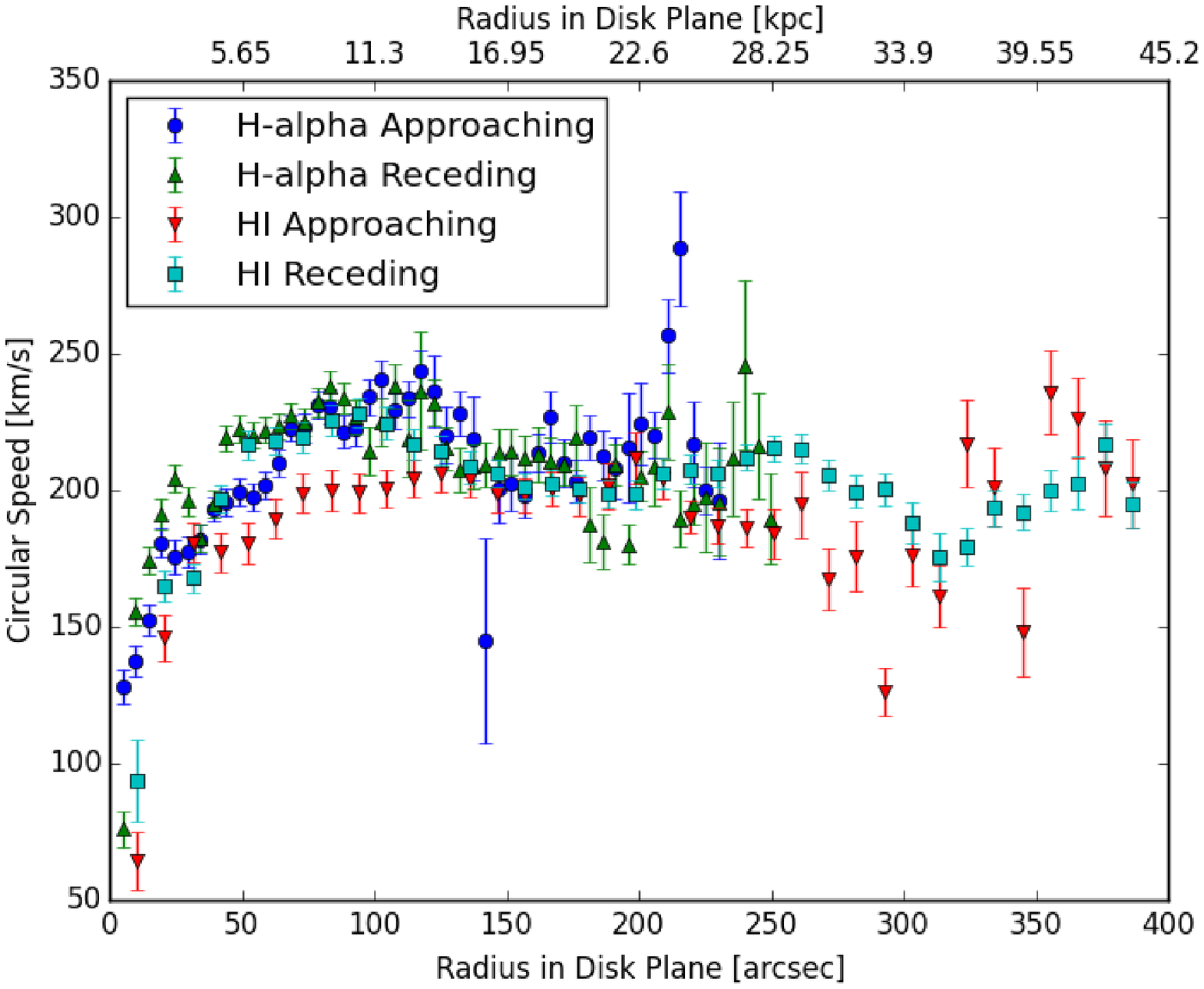}
    \caption{Rotation curves produced from our best-fitting \DF models to the approaching and receding sides of our \HA and \HI velocity maps. Note that the approaching (NW) \HI curve appears to deviate from the other three rotation curves over the region 40\arcsec$<R<120$\arcsec. }\label{fig:splitcurves} \vskip0cm}}

\hbox to \hsize{\box1 \hfil \box2}

\end{figure}

We present the rotation curve derived from our best-fitting \DF model in figure \refnew{fig:rotcurves}.

\section{Comparison to HI Data}

We have also obtained \HI 21 cm aperture synthesis observations of NGC 2280 from the VLA. We present our \HI velocity map in the right panel of figure \refnew{fig:velmaps}. These data are complementary to our \HA \FP data in numerous ways. For example, our \HA data have higher spatial resolution but lower spectral resolution than do our \HI data. The \HI data have a larger filling fraction and spatial extent than do the \HA data, except in the central region of the galaxy where the \HI data exhibits a hole. \HA and \HI are also produced by different phases of the ISM (excited and neutral gas, respectively).

\begin{figure}

\setbox1=\vbox{{\hsize=.45\hsize
    \includegraphics[width=\hsize]{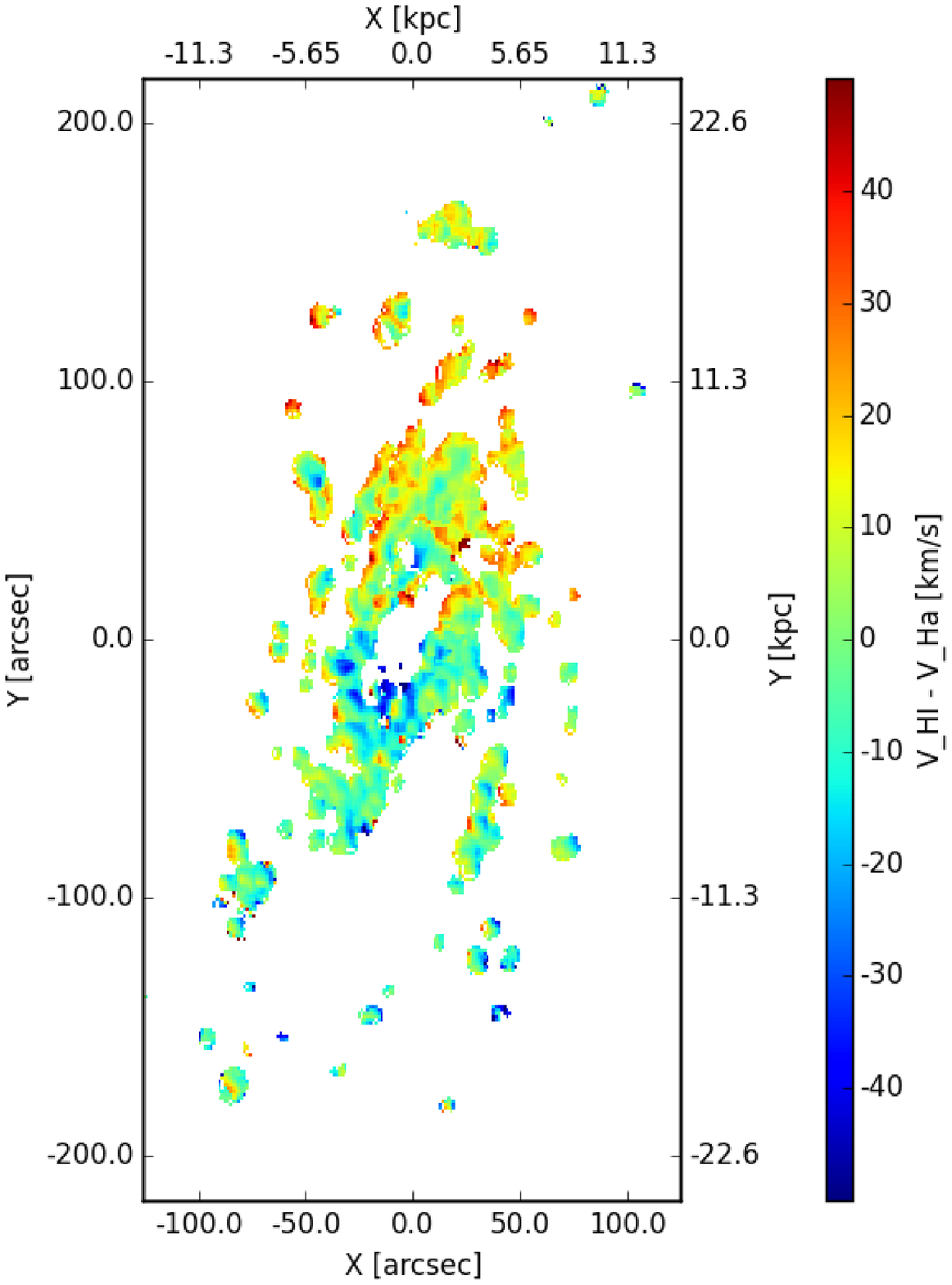}
    \caption{Differences in line-of-sight velocity between our \HA and \HI velocity maps at all points where velocities were measured for both lines.}
\label{fig:veldiff9}
\vskip0.7cm}}

\setbox2=\vbox{{\hsize=.45\hsize 
    \includegraphics[width=\hsize]{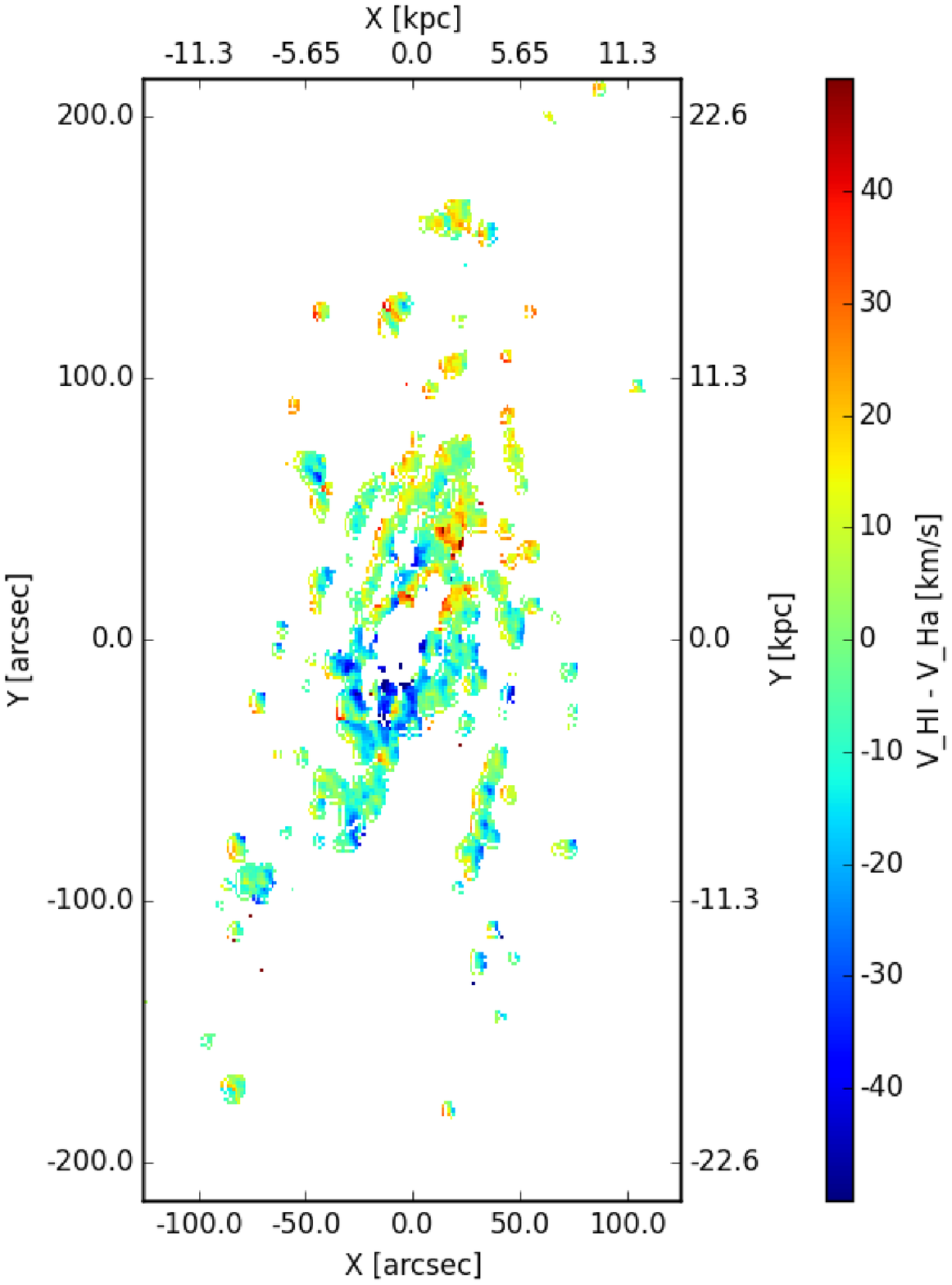}
    \caption{Same as figure 5, except that we have used $5\times5$ pixel binning rather than $9\times9$ in producing our \HA velocity map. Note that some of the regions of large difference near the edges of the map have disappeared, though several areas remain.}
\label{fig:veldiff5}
\vskip0cm}}

\hbox to \hsize{\box1 \hfil \box2}

\end{figure}

\begin{figure}

\setbox1=\vbox{{\hsize=.45\hsize
    \includegraphics[width=\hsize]{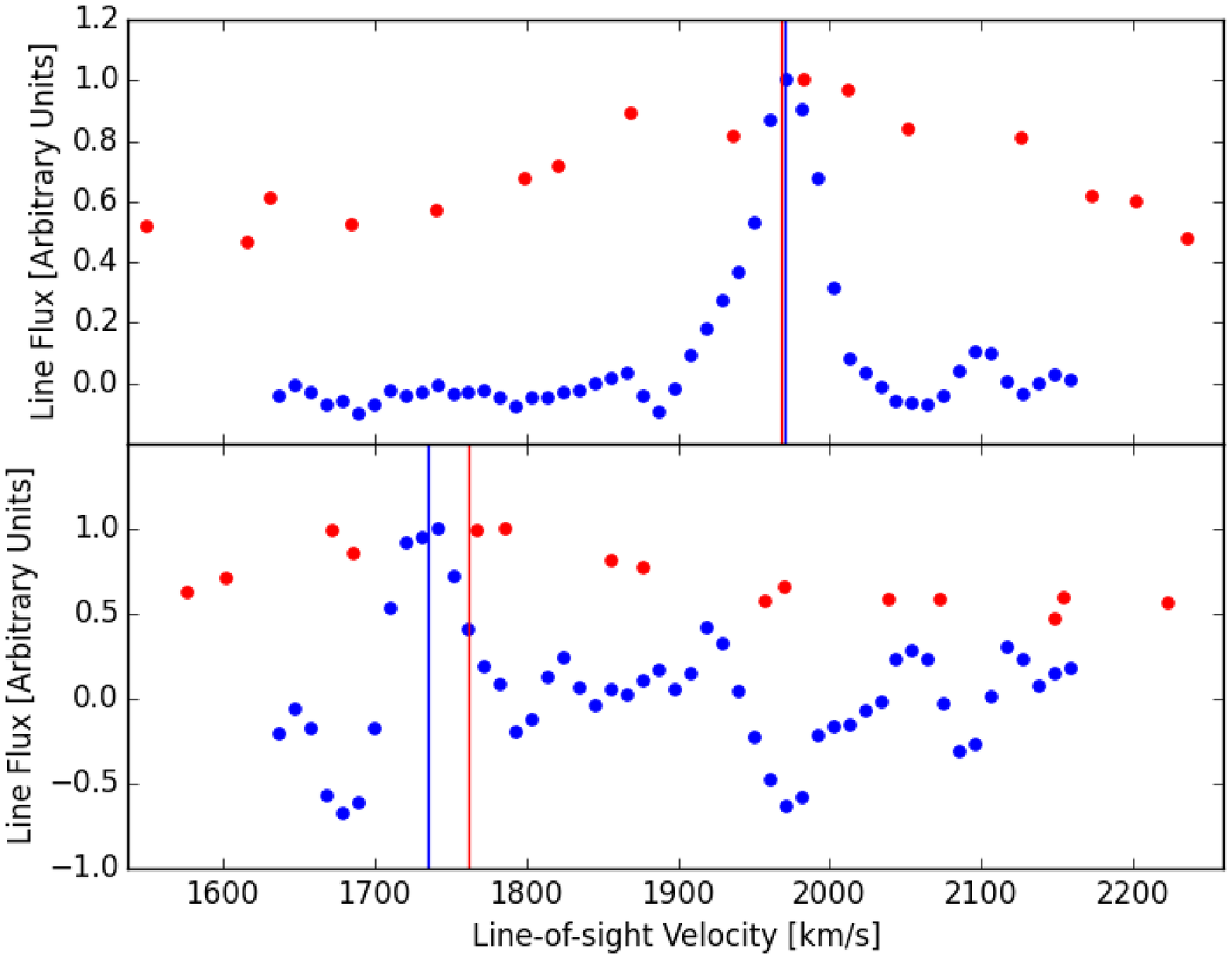}
    \caption{Two sample line profiles comparing our \HA (red) and \HI (blue) data. Note the significantly higher spectral resolution of the \HI spectrum. The profiles in the top panel are typical of our data; the two lines and their fitted velocities agree well. The spectra in the bottom panel correspond to a pixel where the two maps do not agree well; such pixels are rare in our data. Flux units have been rescaled such that all of the spectra peak at a value of 1.}\label{fig:profiles}
\vskip1.6cm}}

\setbox2=\vbox{{\hsize=.45\hsize 
    \includegraphics[width=\hsize]{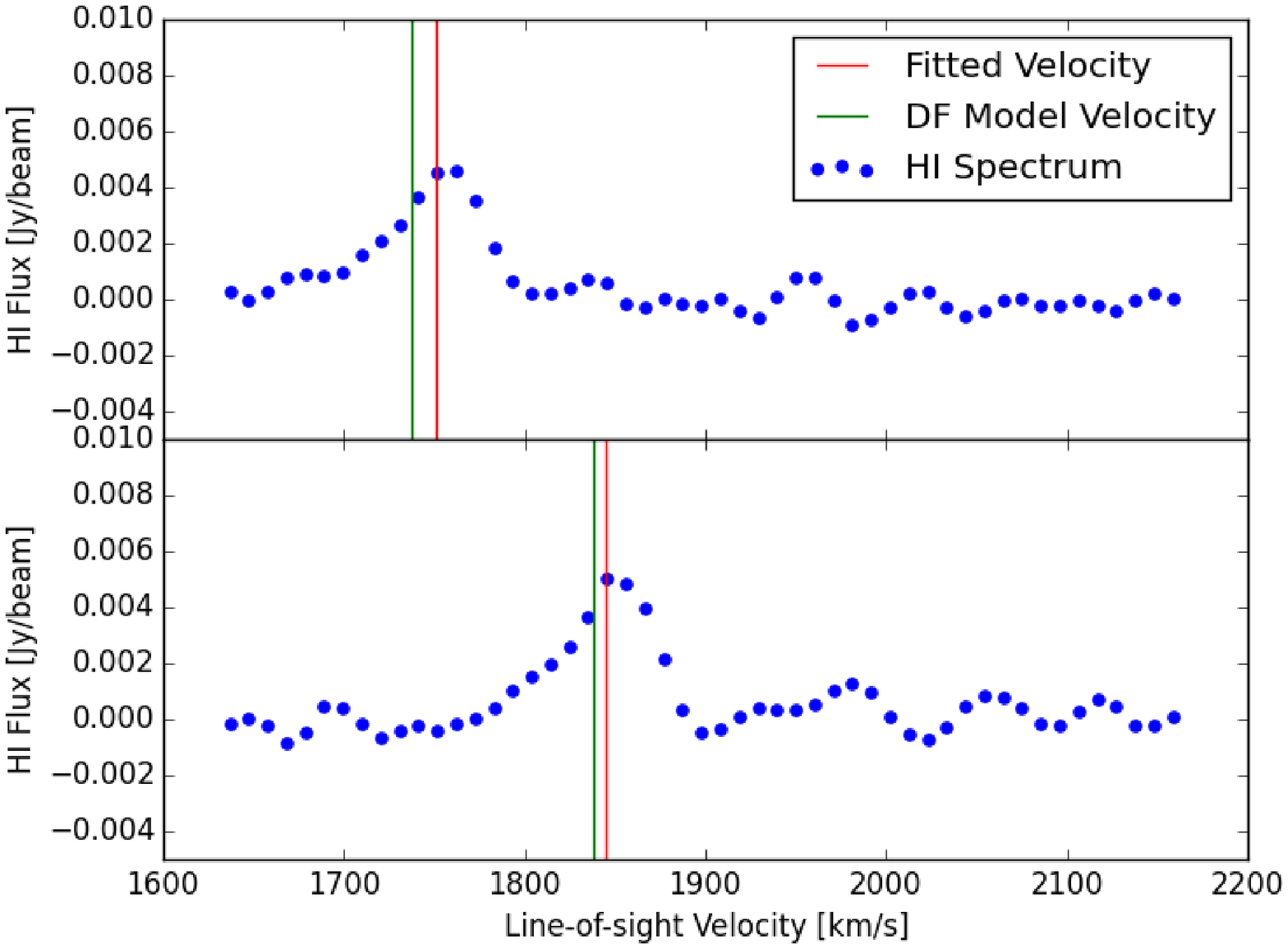}
    \caption{Two sample \HI profiles from the region of large positive residual in the right panel of figure 2. The red line shows the fitted velocity of the \HI spectrum (blue points), while the green line denotes the velocity predicted from our best-fitting \DF models to the entire velocity map. Note the skewness of the profiles, with a ``shoulder'' at lower line-of-sight velocity. We interpret this shoulder as corresponding to the galaxy's rotation and the peak as a stream of gas external to the galaxy. The two spectra shown are from pixels separated by several beam-widths.}\label{fig:h1profiles} \vskip0cm}}

\hbox to \hsize{\box1 \hfil \box2}

\end{figure}

As with our \HA data, we have modeled our \HI data with \DF. The best-fitting systemic velocity and geometric projection parameters agree well both with the same parameters for our models of the \HA \FP data and with previously published values in the literature \citep[see table 3 of][]{Mitc15}. The right panel of figure \refnew{fig:resmaps} shows the data-minus-model residual map for our best fitting \DF model to these data.

We present the rotation curve derived from our best-fitting \DF model to our \HI data in figure \refnew{fig:rotcurves}. At most radii, the two rotation curves agree within their respective error bars. In the region 40\arcsec $< r <$ 120\arcsec the \HA data appear to have a systematically higher circular speed than do the \HI data.

To further investigate this discrepancy, we have produced a map of the difference between the \HA and \HI velocity fields, shown in figure \refnew{fig:veldiff9}. Most of the plotted pixels have values near 0 km/s, demonstrating the close agreement between the two sets of data. However, some spatially coherent regions of large velocity difference exist. Such differences between \HA and \HI kinematic measurements are not a new phenomenon \citep{Phoo93,Zanm08}.

Some of the regions of large velocity difference, in particular those near the edges of the \HA velocity map, may be a result of our choice of 9$\times$9 spatial pixel binning in the \HA data cube. To investigate this, we have reproduced this velocity difference map using 5$\times$5 pixel binning of the \HA data. Figure \refnew{fig:veldiff5} shows the resulting velocity difference map. Many of the regions of large velocity difference are less-pronounced than in the 9$\times$9 binned version of the same figure. Our understanding of this phenomenon is as follows: in spatial regions where line strength is roughly uniform, beam smearing has the effect of blurring velocities equally in all directions. In regions where line strength changes substantially over a small area, a beam placed over this area will blur velocities preferentially in one direction (from high-intensity areas to lower-intensity areas, but not the reverse). The edges of this velocity difference map correspond to a region where the \HI intensity is roughly uniform, but the \HA intensity is not. The fitted \HA velocities do not represent the true velocity at these pixels, and therefore are the cause of the large differences. This effect explains some, but certainly not all, of the features in the velocity difference map.

To better understand these velocity differences, we have examined the line profiles of several individual pixels in our \HI data cube and compared them to the line profiles of the corresponding pixels in our \HA \FP data cube. For the vast majority of pixels, the two spectra and our fits to the line profile agree quite well. We show two sample line profiles in figure \refnew{fig:profiles}. The profiles in the top panel are typical of pixels in our maps, and most of the individual line profiles we have examined are qualitatively similar. The second shows a pixel in which the fitted \HA and \HI velocities differ significantly, taken from a region of high velocity difference in figure \refnew{fig:veldiff9}. These regions have angular sizes comparable to that of our effective \HA beam, indicating they are likely caused by some feature in the \HA velocity field and not the \HI. We believe these features may be evidence of bubbles or chimneys of excited gas in regions of high star formation.

These individual patches of large velocity difference alone do not explain the systematic offset in the \HA and \HI model rotation curves over the region 40\arcsec $< r <$ 120\arcsec. Furthermore, we have fitted \DF models to the approaching and receding halves of the two velocity fields independently. We show the resulting rotation curves in figure \refnew{fig:splitcurves}. Note that the approaching (NW) \HA, receding (SE) \HA, and receding (SE) \HI curves all appear to be consistent with each other over the region of interest, while the approaching (NW) \HI curve appears to be significantly different.

Here we note the presence of a large, spatially coherent, positive residual velocity in the \HI field. This region extends from approximately $(X,Y) = (50$\arcsec$, 100$\arcsec$)$ to $(X,Y) = (0$\arcsec$, -75$\arcsec$)$ in the right panel of figure \refnew{fig:resmaps} and intersects the region where the rotation curve discrepancy is most pronounced. We have examined the \HI line profiles in this region and have found that many of them have a skewed shape or are multiply peaked. We plot two such line profiles in figure \refnew{fig:h1profiles}. The peaks of these profiles lie at higher recessional velocities than are predicted by our best-fitting \DF model. We believe the lower-velocity ``shoulders'' of these asymmetric line profiles correspond to the galaxy's rotation, while the higher-velocity peak corresponds to a stream of neutral gas external to the galaxy. Assuming this stream of gas is in the foreground of NGC 2280, it would correspond to an inflow of neutral gas. We believe this stream of gas may be having a significant effect on the \DF model circular velocity over this radial region, causing the large difference in rotation curves that we observe.

\section{Summary}

By utilizing the SALT \FP interferometer's large field of view and high spatial resolution, we are measuring galaxy velocity fields on previously unmatched scales. Our measurements and subsequent kinematic modeling of the \HA velocity field in NGC 2280 agree very well with both our \HI 21 cm observations and previous measurements in the literature. SALT \FP data for 12 of the 19 galaxies in the RINGS sample are now in hand, and our velocity fields produced from these data in many cases have higher spatial resolutions than that of NGC 2280. These high-resolution kinematic maps are allowing us to probe the structure of these galaxies on smaller scales than previously possible.

\end{document}